\documentstyle[aps,twocolumn,floats]{revtex}
\input psfig.sty
\begin{document}
\draft
\title{
Magnetic Excitations in quasi two--dimensional Spin--Peierls Systems
}
\author{Wolfram Brenig}
\address{Institut f\"ur Theoretische Physik, Universit\"at zu K\"oln
Z\"ulpicher Str. 77, 50937 K\"oln, Germany}
\date{\today}
\maketitle
\begin{abstract}
A study is presented of a two--dimensional frustrated and dimerized
quantum spin--system which models the effect of inter--chain coupling
in a spin--Peierls compound. Employing a bond--boson method to account
for quantum disorder in the ground state the elementary excitations
are evaluated in terms of gapful triplet modes.  Results for the
ground state energy and the spin gap are discussed.  The triplet
dispersion is found to be in excellent agreement with inelastic
neutron scattering data in the dimerized phase of the spin--Peierls
compound CuGeO$_3$. Moreover, consistent with these neutron scattering
experiments, the low--temperature dynamic structure factor exhibits
a high--energy continuum split off from the elementary triplet mode.
\end{abstract}

\section{Introduction}

The recent discovery of the anorganic spin--Peierls compounds
CuGeO$_3$ \cite{Hase93} and $\alpha'$--NaV$_2$O$_5$
\cite{Isobe96a,Weiden97a} has great\-ly stimulated interest in
low--dimensional magne\-tism. While many properties of these novel
materials can be described in terms of quasi {\em one}--dimensional
(1D) quantum spin--systems, clear evidence for a substantial degree of
{\em two}--dimensionality (2D) of their magnetism has been found -- most
noteworthy by inelastic neutron scattering (INS) which displays a
sizeable transverse dispersion of the magnetic excitations in
CuGeO$_3$ \cite{Nishi94,Regnault95a,Regnault96a}.  In the present
study I will establish a simple framework to interpret the spin
dynamics of a frustrated and dimerized 2D quantum spin--model with a
particular focus on the low--temperature phase of CuGeO$_3$.

CuGeO$_3$ is an anorganic spin--Peierls system with a lattice
dimerization transition at a temperature $T_{SP}\simeq 14K$
\cite{Hase93,Nishi94,Regnault95a,Regnault96a,Martin96a,Pouget94}. Its
structure comprises of weakly coupled CuO$_2$ chains along the
c--axis, with copper in a spin--1/2 state
\cite{Voellenkle67,Hirota94a}. The nearest--neighbor ($n.n.$)
exchange--coupling between copper spins along the CuO$_2$ chains is
strongly reduced by almost orthogonal intermediate oxygen states
\cite{Braden96}. Therefore, next--nearest--neighbor ($n.n.n.$)
exchange in CuGeO$_3$ is relevant.  Both, $n.n.$ and $n.n.n.$
exchange, are antiferromagnetic (AFM) \cite{Braden96,Geertsma96}
implying intra--chain frustration. In addition, $n.n.$, as well as
$n.n.n.$ {\em inter}--chain exchange is present which proceeds via the
O2 sites \cite{Braden96,Khomskii96a}. This exchange is believed to be
one order of magnitude less than the intra--chain coupling
\cite{Nishi94,Regnault95a,Regnault96a,Braden96,Khomskii96a} and
comparable to $T_{SP}$. Therefore the inter--chain coupling should be
relevant in the dimerized  phase. This may be a key element in the INS
\cite{Nishi94,Regnault95a,Regnault96a,Martin96a} and magnetic Raman
scattering \cite{Kuroe94,Loosdrecht96,Lemmens96} data.

\begin{figure}[tb]
\vskip .2cm 
\centerline{\psfig{file=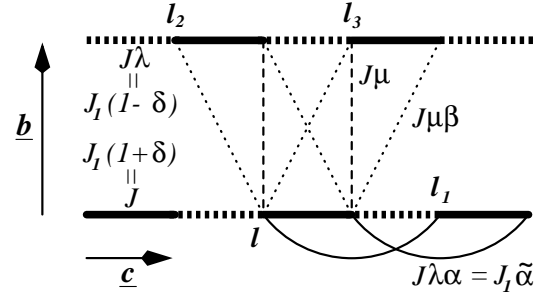,width=7cm}}
\vskip .1cm 
\caption[l]{The $J$-$\lambda$-$\alpha$-$\mu$-$\beta$ model.  Line
segments refer to exchange couplings for spins located at segment
vertices. Inter--chain couplings are shown for a single dimer site
only. ${\bf b}=b{\bf e}_b$ and ${\bf c}=c{\bf e}_c$ are the primitive
vectors.}
\label{bobofig1}
\end{figure}

A minimal model of CuGeO$_3$ which includes intra--, as well as
inter--chain interactions is the $J$-$\lambda$-$\alpha$-$\mu$-$\beta$
model \cite{Braden96,Uhrig97a} depicted in fig.~\ref{bobofig1}. The
various line segments label the coupling strengths
$J,J\lambda,J\lambda\alpha,J\mu$, and $J\mu\beta$ between the spins
located at the vertices in this figure. $J$ refers to the 'strongest'
or dimer--bond the left vertices of which form the dimer lattice ${\bf
l}\in {\cal D}$. Most important the dimerization in
fig.~\ref{bobofig1} is staggered along the b--axis. This is realized
both, in CuGeO$_3$ \cite{Hirota94a} as well as in
$\alpha'$--NaV$_2$O$_5$ \cite{Fujii96a}, and turns out to be relevant
for the magnon dispersion.  In fig.~\ref{bobofig1} an additional,
so--called 'natural' labeling of the intra--chain parameters is
introduced, i.e $J_1,\tilde{\alpha}$, and $\delta$. This notation is
frequently used in the context of the 1D dimerized and frustrated
spin--chain limit.  In CuGeO$_3$ $J_1$ is approximately $160K$
\cite{Castilla95,Riera95}. Consensus on the precise magnitude of the
intra--chain frustration--ratio $\tilde{\alpha}$ is still lacking.
Studies of the magnetic susceptibility, which has been compared only to
1D models, have resulted in $\tilde{\alpha}\approx 0.24$
\cite{Castilla95} as well as in $\tilde{\alpha}\approx 0.35$
\cite{Riera95}. This would place CuGeO$_3$ in the vicinity of the
critical value $\tilde{\alpha}_c \simeq 0.2411$ for the opening of a
spin gap solely due to frustration \cite{Okamoto92a}. $\delta$
resembles the lattice dimerization which is finite for $T<T_{SP}$
only. Values for the zero--temperature dimerization $\delta(T=0)$
ranging from $0.21$ to $0.012$ have been suggested
\cite{Braden96,Castilla95,Riera95,otherdeltas}.  Knowledge on the
magnitude of $\mu$ and $\mu\beta$ is limited to $|\mu|,|\mu\beta|\ll 1$
\cite{Braden96,Khomskii96a}.

Magnetic excitations in CuGeO$_3$ are clearly distinct among the
uniform (U), i.e. $T>T_{SP}$, and the dimerized (D), i.e. $T<T_{SP}$,
phase. While the dynamic structure factor exhibits a gapless, c--axis
dispersive two--spinon continuum similar to that of the 1D Heisenberg
chain above $T_{SP}$ \cite{Arai96}, well defined magnon--like
excitations with sizeable c-- and b--axis dispersion have been
observed below $T_{SP}$ \cite{Nishi94,Regnault95a,Regnault96a,Ain97a}.
These magnons are gapful and are split off from a continuum which, at
zone--center, starts at roughly twice the magnon gap
\cite{Ain97a,Uhrig96a,Fledderjohann96,Uhrig97a}. 

The aim of this work is to study the magnetic properties of the
$J$-$\lambda$-$\alpha$-$\mu$-$\beta$ model with a focus on the
D--phase of CuGeO$_3$.  First I will describe a bond--spin
representation of the $J$-$\lambda$-$\alpha$-$\mu$-$\beta$ model
which, in turn, is treated by an appropriate linearization. Next,
results for the ground state energy, the spin gap, the magnon
dispersion, and the dynamic structure factor are contrasted against
other theoretical approaches and are compared with experimental
findings. Finally, details of an alternative mean--field approach
using the bond--spin representation are provided in appendix \ref{A}.

\section{Bond--Operator Theory}

In this section the properties of the
$J$-$\lambda$-$\alpha$-$\mu$-$\beta$ model are discussed by
representing the {\em site}--spin algebra in terms of {\em
bond}--spin operators \cite{Sachdev90a}. First, the essential features
of these operators are briefly restated. Consider any two spin--1/2
operators ${\bf S}_1$ and ${\bf S}_2$. The eigenstates of the related
total spin are a singlet $\left|s\right>$ and three triplets
$\left| t_\alpha^{\phantom{\dagger}} \right>$ with $\alpha= x,y,z$.
These can be created out of a vacuum $\left|0\right>$ by
applying the bosonic operators $s_{\phantom{x}}^\dagger$ and
$t_\alpha^\dagger$
\begin{equation}\label{1}
\begin{array}{lccr}
s_{\phantom{.}}^\dagger\left|0\right>
&\hat{=}&\left|s_{\phantom{.}}^{\phantom{\dagger}}\right>=&
(\left|\uparrow\downarrow\right>-\left|\downarrow\uparrow\right>)/
\sqrt{2}\\
t_x^\dagger \left|0\right>
&\hat{=}&\left|t_x^{\phantom{\dagger}}\right>=&
-(\left|\uparrow\uparrow\right>-\left|\downarrow\downarrow\right>)/
\sqrt{2}\\
t_y^\dagger \left|0\right>
&\hat{=}&\left|t_y^{\phantom{\dagger}}\right>=&
i(\left|\uparrow\uparrow\right>+\left|\downarrow\downarrow\right>)/
\sqrt{2}\\
t_z^\dagger \left|0\right>
&\hat{=}&\left|t_z^{\phantom{\dagger}}\right>=&
(\left|\uparrow\downarrow\right>+\left|\downarrow\uparrow\right>)/
\sqrt{2}
\end{array}
\;\;\;,\end{equation}
where $[s,s_{\phantom{\alpha}}^\dagger]=1$,
$[s_{\phantom{\alpha}}^{(\dagger)},t_\alpha^{(\dagger)}]=0$, and
$[t_\alpha^{\phantom{(\dagger)}},t_\beta^{\dagger}]=\delta_{\alpha
\beta}$.
The action of ${\bf S}_1$ and ${\bf S}_2$ on this space leads
to the representation
\begin{equation}\label{2}
S^\alpha_{\stackrel{{\scriptstyle 1}}{2}}\hat{=}
\frac{1}{2} ( \pm s_{\phantom{\alpha}}^\dagger t_\alpha^{\phantom{
\dagger}}
\pm t_\alpha^\dagger s
- i \varepsilon_{\alpha\beta\gamma} t_\beta^\dagger
t_\gamma^{\phantom{\dagger}} )
\;\;\;,\end{equation}
for the individual spin operators. Here
$\varepsilon_{\alpha\beta\gamma}$ is the Levi--Civita symbol and a
summation over repeated
indices is implied hereafter.
The upper(lower) subscript on the lhs. of (\ref{2}) refer to
upper(lower) sign on the rhs..  The bosonic Hilbert space
has to be restricted to the physical Hilbert space, i.e.
either one singlet or one triplet, by the constraint
\begin{equation}\label{3}
s_{\phantom{\alpha}}^{\dagger}s +
t_\alpha^{\dagger}t_\alpha^{\phantom{\dagger}} =1
\;\;\;.\end{equation}
Using (\ref{2}) and (\ref{3}) it is simple to check, that ${\bf S}_1$
and ${\bf S}_2$ satisfy a spin algebra indeed and moreover that
\begin{equation}\label{4}
S^\alpha_1 S^\alpha_2 = -\frac{3}{4} s_{\phantom{\alpha}}^{\dagger}s
+ \frac{1}{4} t_\alpha^{\dagger}t_\alpha^{\phantom{\dagger}}
\;\;\;.\end{equation}

In order to transform the $J$-$\lambda$-$\alpha$-$\mu$-$\beta$ model
into the boson representation a particular distribution '${\bf l}$' of
bonds, i.e. pairs of spins ${\bf S}_{{\bf l}\,1}$ and ${\bf S}_{{\bf
l}\,2}$, has to be selected. Here this selection will be based on the
limit of strong dimerization $(\lambda,\lambda\alpha)\rightarrow
(0,0)$, or equivalently $(\tilde{\alpha},\delta)\rightarrow (0,1)$,
and small inter--chain coupling $(\mu,\mu\beta)\rightarrow (0,0)$. In
this limit the ground state is a product of singlets on each
dimer--bond while the elementary excitations are composed of the
corresponding localized triplets. Therefore it is natural to place the
singlet and triplet bosons onto the dimer bonds, i.e. ${\bf S}_{{\bf
l}\,1}={\bf S}_{{\bf l}}$ and ${\bf S}_{{\bf l}\,2}={\bf S}_{{\bf
l+c}}$ with ${\bf l}\in{\cal D}$.  The transformed Hamiltonian reads
\begin{eqnarray}\label{5}
H&=& H_0+H_1+H_2+H_3
\\
H_0&=& \sum_{{\bf l}\in{\cal D}} (
-\frac{3}{4} s_{{\bf l}}^{\dagger}
s_{{\bf l}}^{\phantom{\dagger}}
+\frac{1}{4} t_{{\bf l}\,\alpha}^{\dagger}
t_{{\bf l}\,\alpha}^{\phantom{\dagger}})
\nonumber \\
H_1&=& \sum_{{\bf l}\neq {\bf m}\in{\cal D}}
a({\bf l},{\bf m}) (
t_{{\bf l}\,\alpha}^{\dagger}
t_{{\bf m}\,\alpha}^{\phantom{\dagger}}
s_{{\bf m}}^{\dagger}
s_{{\bf l}}^{\phantom{\dagger}} +
t_{{\bf l}\,\alpha}^{\dagger}
t_{{\bf m}\,\alpha}^{\dagger}
s_{{\bf m}}^{\phantom{\dagger}}
s_{{\bf l}}^{\phantom{\dagger}} + h.c.)
\nonumber \\
H_2&=& \sum_{{\bf l}\neq {\bf m}\in{\cal D}}
b({\bf l},{\bf m}) (
i \varepsilon_{\alpha\beta\gamma}
t_{{\bf m}\,\alpha}^{\dagger}
t_{{\bf l}\,\beta}^{\dagger}
t_{{\bf l}\,\gamma}^{\phantom{\dagger}}
s_{{\bf m}}^{\phantom{\dagger}} + h.c.)
\nonumber \\
H_3&=& \sum_{{\bf l}\neq {\bf m}\in{\cal D}}
c({\bf l},{\bf m}) (
t_{{\bf l}\,\alpha}^{\dagger}
t_{{\bf m}\,\alpha}^{\dagger}
t_{{\bf m}\,\beta}^{\phantom{\dagger}}
t_{{\bf l}\,\beta}^{\phantom{\dagger}} -
t_{{\bf l}\,\alpha}^{\dagger}
t_{{\bf m}\,\beta}^{\dagger}
t_{{\bf m}\,\alpha}^{\phantom{\dagger}}
t_{{\bf l}\,\beta}^{\phantom{\dagger}})
\nonumber
\;\;\;,\end{eqnarray}
where each local Hilbert space is subject to the constraint (\ref{3})
and, if not explicitly stated otherwise, the unit of energy is $J$
hereafter. The inter--dimer matrix elements can be obtained from
fig.~\ref{1}
\begin{eqnarray}\label{6}
&&a({\bf l},{\bf m}) =-\frac{1}{4}[
t_1 \delta_{{\bf m}{\bf l}_{\scriptstyle 1}} +
t_2 (\delta_{{\bf m}{\bf l}_{\scriptstyle 2}}+\delta_{{\bf m}
{\bf l}_{\scriptstyle 3}})]
\\
&&b({\bf l},{\bf m}) =
\frac{1}{4}[\lambda
(\delta_{{\bf l}{\bf l}_{\scriptstyle 1}}-\delta_{{\bf m}{\bf l}_{
\scriptstyle 1}}) +
\mu (\delta_{{\bf m}{\bf l}_{\scriptstyle 3}}
-\delta_{{\bf m}{\bf l}_{\scriptstyle 2}}
+\delta_{{\bf l}{\bf l}_{\scriptstyle 2}}
-\delta_{{\bf l}{\bf l}_{\scriptstyle 3}})]
\nonumber \\
&&c({\bf l},{\bf m}) =
-\frac{1}{4}[
t_3 \delta_{{\bf m}{\bf l}_{\scriptstyle 1}} +
t_4 (\delta_{{\bf m}{\bf l}_{\scriptstyle 2}}+\delta_{{\bf m}
{\bf l}_{\scriptstyle 3}})]
\nonumber
\;\;\;,\end{eqnarray}
where ${\bf l}_{1,2,3}$ are defined in fig.~\ref{bobofig1} and $t_1=
\lambda(1-2\alpha)$, $t_2=\mu (1-2\beta)$, $t_3=\lambda (1+2\alpha)$, and
$t_4=\mu (1+2\beta)$. As anticipated, the inter--dimer matrix elements
$a({\bf l},{\bf m})$, $b({\bf l},{\bf m})$, and $c({\bf l}, {\bf m})$
vanish in the strong dimer limit leaving the Hamiltonian diagonal in
$s_{{\bf l}}$ and $t_{{\bf l}\,\alpha}$.

In order to treat the local constraint and the dimer interactions
approximations have to be made.  To this end I will employ the
Holstein--Primakoff (HP) representation of the bond--operators which
has been detailed in \cite{Starykh96a,Chubokov89a,Chubukov91a}. In
this representation the constraint is treated by eliminating the
singlet operator via $s_{{\bf l}}^{\dagger}=s_{{\bf
l}}^{\phantom{\dagger}}=(1-t_{{\bf l}\,\alpha}^{ \dagger}t_{{\bf
l}\,\alpha}^{\phantom{\dagger}})^{-1/2}$. Moreover, after inserting
this into the Hamiltonian only terms up to second order in in the
triplet operators are retained. The latter procedure is analogous to
the linear spin--wave approximation in systems with broken
spin--rotational invariance. The linearized (LHP) Hamiltonian is
given by
\begin{eqnarray}\label{7}
H_{LHP}&&=-\frac{9}{4}D
\nonumber \\
&&+\frac{1}{2}\;\sum_{{\bf k}\in{\cal B}}
\Psi_{{\bf k}\,\alpha}^\dagger
\left[\begin{array}{cc}
1+\epsilon_{\bf k} & \epsilon_{\bf k} \\
\epsilon_{\bf k} & 1+\epsilon_{\bf k}
\end{array}\right]
\Psi_{{\bf k}\,\alpha}^{\phantom{\dagger}}
\\ \nonumber \\ \label{7a}
\epsilon_{\bf k}=&&-\frac{1}{2}[t_1 \cos(2k_c)
+2t_2\cos(k_b)\cos(k_c)]
\;\;\;,\end{eqnarray}
where $D$ is the number of dimers and ${\bf k}$ is a momentum
vector restricted to a Wigner--Seitz
cell ${\cal B}$ of the reciprocal lattice. For comparison with
experimental data ${\cal B}$ is oriented with respect to the
non--dimerized system, instead of the Brillouin zone of the dimer
lattice, i.e.  ${\bf k}=(k_b,k_c)$ with $bk_b=0...2\pi$ and
$ck_c=0...\pi$ with $b,c$ set to unity hereafter. $\Psi_{{\bf
k}\,\alpha}^{(\dagger)}$ is a a spinor with $\Psi_{{\bf
k}\,\alpha}^\dagger=[t_{{\bf k}\,\alpha}^{\dagger} \; t_{-{\bf
k}\,\alpha}^{\phantom{\dagger}}]$ and $t_{{\bf l}\,\alpha
}^\dagger=1/\sqrt{D}\sum_{\bf k}e^{-i{\bf k}\cdot{\bf l}}t_{{\bf k}\,
\alpha}^\dagger$.

\begin{figure}[tb]
\vskip -3.3cm 
\centerline{\hspace*{1cm}\psfig{file=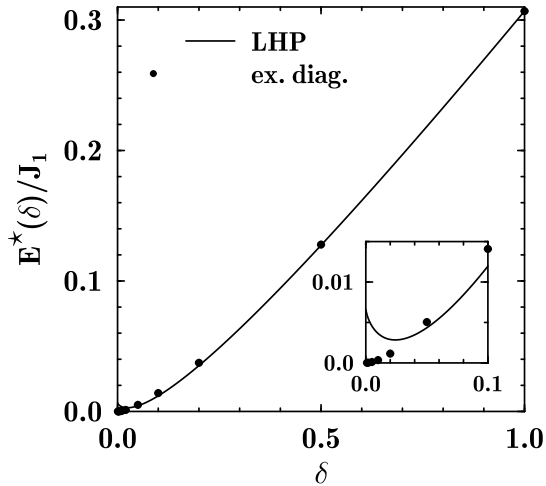,width=9.5cm}}
\vskip -3.3cm 
\caption[l]{\begin{sloppypar}
Ground state energies in the dimerized--chain limit: LHP--theory
(solid) versus exact diagonalization \cite{Soos85a} (solid dots),
errors are less than marker size. Inset: small $\delta$
limit. $E^{\star}(\delta)/J_1=$ $-[(1+\delta) E_g((1-\delta)
/(1+\delta),0)-$ $E^{Bethe}_g]$.
\end{sloppypar}}
\label{bobofig2}
\end{figure}

Eqn. (\ref{7}) describes a threefold degenerate set of dispersive
triplets.  The triplets are renormalized by ground--state
quantum--fluctuations.  These are produced by the terms of type
$t_{{\bf k}\,\alpha}^\dagger t_{-{\bf k}\,\alpha}^\dagger$ and their
hermitian conjugate. The excitation spectrum $E_{\bf k}$ follows from
a Bogoliubov transformation
\begin{eqnarray}\label{8}
&&H_{LHP}=-\frac{9}{4}D+
\sum_{{\bf k}\in{\cal B},\alpha}
E_{\bf k}(a_{{\bf k}\,\alpha}^{\dagger}
a_{{\bf k}\,\alpha}^{\phantom{\dagger}}
+\frac{1}{2})
\\ \label{8a}
&&E_{\bf k}=\sqrt{1+2\epsilon_{\bf k}}
\;\;\;,\end{eqnarray}
where $a_{{\bf k}\,\alpha}^{(\dagger)}$ are the Bogoliubov
quasi--particles which are given by
\begin{eqnarray}\label{9}
&&\Psi_{{\bf k}\,\alpha}^{\phantom{\dagger}}=
\left[\begin{array}{cc}
g_{\bf k} & h_{\bf k} \\
h_{\bf k} & g_{\bf k}
\end{array}\right]
\Phi_{{\bf k}\,\alpha}^{\phantom{\dagger}}
\\  \nonumber \\
&&h^2_{\bf k}=
\frac{1}{2}\left(\frac{1+\epsilon_{\bf k}}{E_{\bf k}}-1\right)
\makebox[1cm][c]{;}
h_{\bf k}g_{\bf k}=-\frac{1}{2}\frac{\epsilon_{\bf k}}{E_{\bf k}}
\nonumber \\
&&g^2_{\bf k}=
\frac{1}{2}\left(\frac{1+\epsilon_{\bf k}}{E_{\bf k}}+1\right)
\nonumber
\;\;\;,\end{eqnarray}
where $\Phi_{{\bf k}\,\alpha}^{(\dagger)}$ is a a spinor with
$\Phi_{{\bf k}\,\alpha}^\dagger=[a_{{\bf k}\,\alpha}^{ \dagger} \;
a_{-{\bf k}\,\alpha}^{\phantom{\dagger}}]$.

To conclude this section, I note that instead of treating the
constraint by means of the HP representation one may also apply the so
called bond--operator mean--field theory (MFT) of
\cite{Sachdev90a}. The application of this method to the
$J$-$\lambda$-$\alpha$-$\mu$-$\beta$ model
is detailed in appendix \ref{A}. As will become evident in section
\ref{GapSection} this technique seems less well suited in the present
context.

\begin{figure}[tb]
\vskip -3.3cm 
\centerline{\hspace*{.2cm}\psfig{file=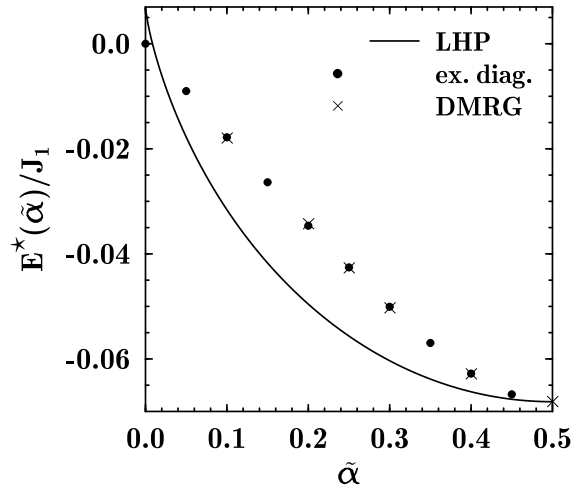,width=9.5cm}}
\vskip -3.3cm 
\caption[l]{\begin{sloppypar}
Ground state energies in the frustrated--chain limit:
LHP--theory (solid) versus exact diagonalization \cite{Tonegawa87a}
(solid dots), and DMRG \cite{Chitra95a} (crosses), errors are less
than marker size. $E^{\star}(\alpha)/J_1=$ $-[E_g((1-2\alpha),0)-$
$E^{Bethe}_g]$.
\end{sloppypar}}
\label{bobofig3}
\end{figure}

\section{Results}

In the following sections the consequences of the LHP representation
of the 2D $J$-$\lambda$-$\alpha$-$\mu$-$\beta$ model will be
contrasted against other known results in the limiting case of the
1D $J_1$-$\tilde{\alpha}$-$\delta$ model as well as INS data observed
on CuGeO$_3$.

\subsection{Ground State Energy}

From (\ref{8}) it is obvious, that the ground state energy per {\em
lattice site}, $E_g$, is equal to $-3/8$ at the point of complete
dimerization. This is the proper energy gain for a bare singlet
formation. For arbitrary $t_1$ and $t_2$
\begin{eqnarray}\label{15}
\lefteqn{
E_g(t_1,t_2)=-\frac{9}{8}+\frac{3}{2\pi^2}\int_{0}^{\pi}dk_c\,
\{[1 - 2 t_2 \cos(k_c) }
\\
&& - t_1 \cos(2 k_c)]^{-1/2} \mbox{{\bf E}}\,(
\frac{-4 t_2 \cos(k_c)}{1 - 2 t_2 \cos(k_c) - t_1 \cos(2 k_c)}) \}
\nonumber
\;\;\;,\end{eqnarray}
where $\mbox{{\bf E}}$ is the complete elliptic integral of the second
kind.  For vanishing inter--chain coupling this simplifies to
\begin{equation}\label{16}
E_g(t_1,0)= -9/8 + \frac{3}{2\pi} \sqrt{1 - t_1}\mbox{{\bf E}}(\frac{2
t_1}{t_1-1})
\;\;\;,\end{equation}
which can be compared to existing results in various regions of the
$(\tilde{\alpha},\delta)$--plane of the frustrated and dimerized
spin--1/2 chain. In particular at the isotropic Heisenberg point,
i.e. $(\tilde{\alpha},\delta)=(0,0)$, $E_g=3/(\sqrt{2}\pi)-9/8\approx
-0.4498$ which agrees reasonably well with the Bethe Ansatz result
$E^{Bethe}_g=1/4-\ln (2)\approx -0.4431$. In fig.~\ref{bobofig2} the
ground state energy along the $(\tilde{\alpha}=0,\delta)$--line,
i.e. for a dimerized chain, is contrasted against results from exact
diagonalization \cite{Soos85a}. Deviations are within the numerical
error of the diagonalization data for $\delta\gtrsim 0.05$. For
dimerizations below $0.05$ small differences are caused by an
unphysical extremum in the LHP energy which is visible in the
inset. While the agreement is encouraging along the
$(\tilde{\alpha}=0,\delta)$--line, only qualitative consistency is to
be expected along the $(\tilde{\alpha},\delta=0)$--line since this
region is more distant from the strong dimer limit
$(\tilde{\alpha},\delta)\sim (0,1)$. This is shown in
fig.~\ref{bobofig3} which compares $E_g(\tilde{\alpha})$ with exact
diagonalization \cite{Tonegawa87a} and density--matrix renormalization
group (DMRG) \cite{Chitra95a} data for the case of a frustrated
chain. This figure demonstrates that the LHP approach overestimates
the frustration induced loss of ground state energy at intermediate
$\tilde{\alpha}$.

\begin{figure}[tb]
\vskip -3.3cm 
\centerline{\hspace*{1cm}\psfig{file=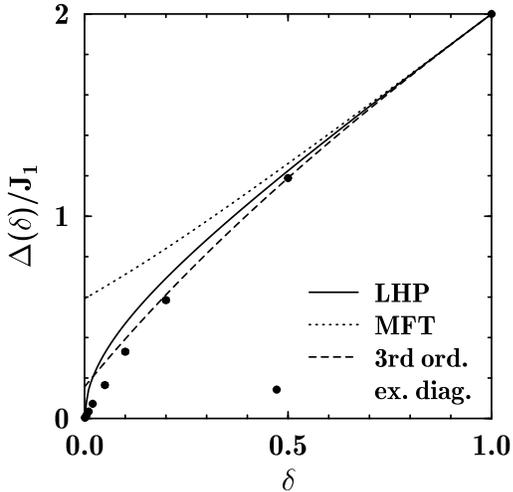,width=9.5cm}}
\vskip -3.3cm 
\caption[l]{\begin{sloppypar}
Spin gaps in the dimerized--chain limit: LHP--\-theory (solid)
versus 3rd--order perturbation theory \cite{Uhrig97a,Harris73a} ( dashed),
MFT (dotted), and exact diagonalization \cite{Soos85a} (solid dots),
errors are less than marker size.
\end{sloppypar}}
\label{bobofig4}
\end{figure}

\begin{figure}[tb]
\vskip -3.3cm 
\centerline{\hspace*{1cm}\psfig{file=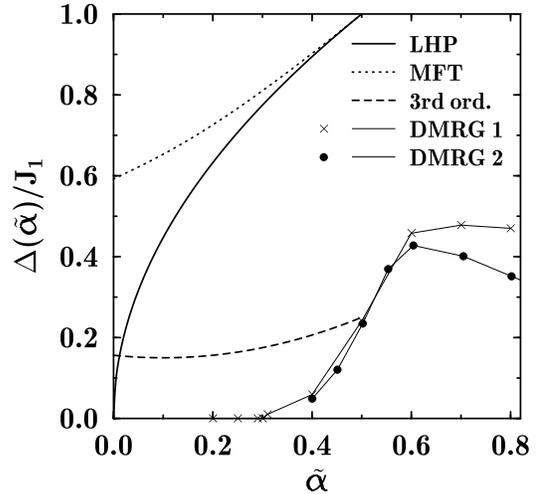,width=9.5cm}}
\vskip -3.3cm 
\caption[l]{\begin{sloppypar}
Spin gaps in the frustrated--chain limit: LHP--\-theory ( solid )
versus 3rd--order perturbation theory \cite{Uhrig97a} (dashed), MFT
(dotted), and two DMRG calculations \cite{Chitra95a} (crosses),
and \cite{White96a} (solid dots), errors are less than marker size.
\end{sloppypar}}
\label{bobofig5}
\end{figure}

\subsection{Spin Gap}\label{GapSection}

The LHP approximation does not break the spin--rotational invariance
and leaves the system in a quan\-tum--disordered ground state. This is
consistent with a spin gap $\Delta$ of the triplet dispersion which,
for positive $t_1$ and $t_2$, is given by $\Delta = \sqrt{1-t_1-2
t_2}$ and is situated at ${\bf k}=(0,0)$ and $(\pi,\pi)$. If $t_1$ and
$t_2$ are such that the triplet modes turn massless, i.e. $\Delta=0$,
a quantum phase--transition towards antiferromagnetism (AFM) will
occur at $T=0$.  In terms of $\tilde{\alpha}$ and $\delta$, the AFM
instability--line is located at $\tilde{\alpha}+\delta=t_2(1+\delta)$.
Beyond this line the LHP approach breaks down.

Next the LHP spin gap at $t_2=0$ is compared to known results for the
frustrated and dimerized spin--1/2 chain. In fig.~\ref{bobofig4}
$\Delta(\delta)$ is contrasted against findings of exact
diagonalization \cite{Soos85a}, perturbation theory up to third order
in $\lambda$ \cite{Uhrig97a,Harris73a}, and a solution of the MFT equations
(\ref{12a}--\ref{13a}) which are discussed in appendix \ref{A}. This
figure displays reasonable agreement between the first three of these
approaches for all values of $\delta$. In addition it shows that the
MFT suffers from the inability to close the spin gap at the isotropic
Heisenberg point \cite{Zang95a,3rdNoZero}. This caveat renders the MFT
unsuitable for the case of a systems with nearly massless spin
excitations, e.g. CuGeO$_3$.

As noted in the previous section, analytic approaches based on the
strong dimer limit are less reliable if considered along the
$(\tilde{\alpha}, \delta=0)$--line. Nevertheless, a comparison of
$\Delta(\tilde{\alpha})$ as obtained from the LHP theory with various
other techniques is instructive and is shown in fig.~\ref{bobofig5}.
Qualitatively, all analytic methods depicted exhibit a tendency of the
spin gap to increase as $\tilde{\alpha}$ increases \cite{3rdHasMin}
however agreement with the DMRG data \cite{Chitra95a,White96a} is absent.
In particular, while 3rd--order perturbation theory and MFT show no
critical frustration--ratio $\tilde{\alpha}_c$ and have a finite gap for
all $0<\tilde{\alpha}<0.5$ the LHP representation leads to a critical
frustration $\tilde{\alpha}_c=0$. Quite remarkably there are also
substantial differences between the two DMRG results for
$\tilde{\alpha}\gtrsim 0.5$ \cite{Karen97a}.

\subsection{Triplet Dispersion}\label{DispSection}

In this section the relevance of the model with respect to the spin
excitations in CuGeO$_3$ will be assessed by comparison of $E_{\bf k}$
with INS data for the D-phase. In particular the role of
two--dimensionality will be considered. The essential effect of a
finite hopping amplitude $t_2$ is a {\em mixing} of the b-- and
c--axis dispersion.  This mixing is due to the staggering of the
dimerization along the b--axis and leads to the
$t_2\cos(k_b)\cos(k_c)$--term in $\epsilon_{\bf k}$. Thus, $E_{\bf k}$
involves terms of different periodicity in $k_c$, i.e. $\cos(2k_c)$
and $\cos(k_c)$. Therefore, in contrast to the quasi 1D case, the
degeneracy of the triplets at the momenta $(0,0)$ and $(0,\pi)$ is
lifted.  An identical reasoning based on the first--order contribution
of a 3rd--order perturbation theory has been given in \cite{Uhrig97a}.
Expanding $E_{\bf k}$ in terms of $t_1$ and $t_2$ I find
agreement up to first order with the dispersion given in
\cite{Uhrig97a}.

\begin{figure}[tb]
\vskip -3.2cm 
\centerline{\psfig{file=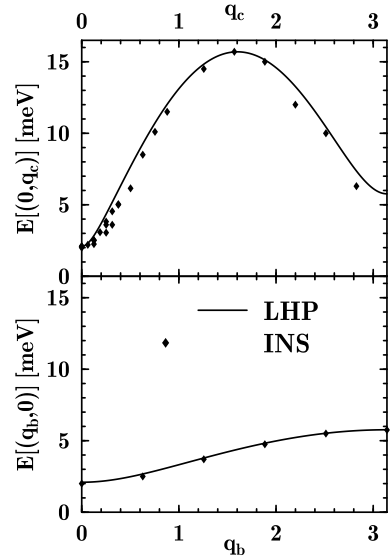,width=9.5cm}}
\vskip -2.4cm 
\caption[l]{\begin{sloppypar}
INS data of c--axis (upper panel) and b--axis (lower panel) dispersion
of the triplet mode in the D--phase of CuGeO$_3$ at $T=1.8K$
\cite{Regnault96a} (solid diamonds) versus LHP--theory (solid) for
$J=11.5meV$, $t_1=0.859$ and $t_2=0.054$. ( $b=c=1$ ).
\end{sloppypar}}
\label{bobofig6}
\end{figure}

In fig.~\ref{bobofig6} INS data of the magnon dispersion in CuGeO$_3$
are shown for momenta along the edges of the reciprocal lattice cell
from ${\bf k}=(0,0)$ to $(0,\pi)$ as well as from $(0,0)$ to
$(\pi/2,0)$ \cite{Regnault96a}. Although data exactly at $(0,\pi)$ is
lacking, it seems very likely from this figure that the triplet
excitations in CuGeO$_3$ are {\em not} degenerate at $(0,0)$ and
$(0,\pi)$. In order to demonstrate that the
$J$-$\lambda$-$\alpha$-$\mu$-$\beta$ model can account for the
observed dispersion fig.~\ref{bobofig6} contains a comparison of
$E_{\bf k}$ with the INS data. Rather than performing this comparison
by a least--square fit, $E_{\bf k}$ has been identified with the
magnon energies only at ${\bf k}=(0,0)$, $(0,\pi/2)$, and
$(\pi/2,0)$. This fixes $J=11.5meV$, $t_1=0.859$, and $t_2=0.054$
unambiguously and limits the b-- to c--axis coupling--ratio to roughly
6\% which is consistent with
\cite{Nishi94,Regnault95a,Regnault96a,Khomskii96a}.  While the
preceding leaves $\tilde{\alpha}$, $\delta$, $\beta$, and $\mu$
undetermined, $\tilde{\alpha}$ can be fixed using an additional input,
i.e. $\delta= 0.012$. This is within the range of values suggested in
the literature, i.e. $\delta=0.21 ... 0.012$
\cite{Braden96,Castilla95,Riera95,otherdeltas} and implies natural
parameters of $J_1=11.4meV=132K$ and $\tilde{\alpha}=0.059$.  Allowing
for a larger dimerization leads to smaller $J_1$ and
$\tilde{\alpha}$. In particular $\tilde{\alpha}=0$ with $J_1=124K$ is
reached for $\delta=0.076$. Parameters with a slightly smaller(larger)
intra(inter)--chain exchange have been established in \cite{Uhrig97a},
i.e. $J_1=9.8meV=114K$, $\tilde{\alpha}=0$, and $t_2=0.12$ however
with $\delta=0.12$.

The agreement displayed in fig.~\ref{bobofig6} is satisfying and
demonstrates the main point of this section, i.e. that diagonal
triplet--hopping can account for the observed asymmetry of the INS
data in the D-phase of CuGeO$_3$. Moreover, the values of
$\tilde{\alpha}$ obtained suggests that the intra--chain frustration
in CuGeO$_3$ is significantly smaller than that derived from the purely
1D $J_1$-$\tilde{\alpha}$-$\delta$ model \cite{Castilla95,Riera95},
i.e. $0.24\lesssim\tilde{\alpha}\lesssim 0.36$. The determination of
model parameters however is in need of further studies to improve on
the quantitative reliability of the frustration--dependence of
presently available approaches for the 2D case.
This is in contrast to \cite{Uhrig97a} where
{\em quantitative} evidence for $\tilde{\alpha}\approx 0$, based on
3rd--order perturbation theory, has been suggested.

\begin{figure}[tb]
\vskip -2.9cm 
\centerline{\psfig{file=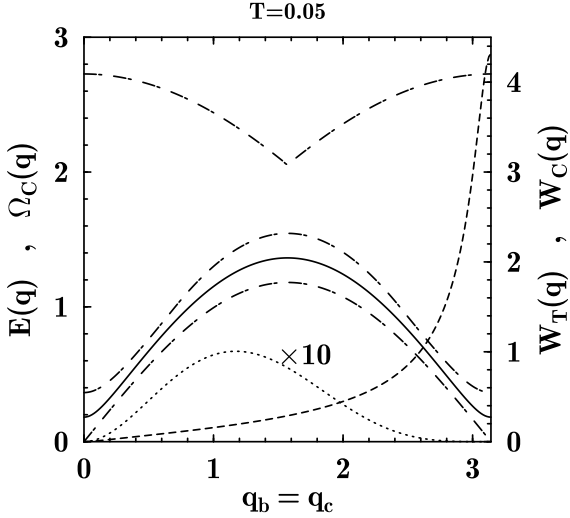,width=9.5cm}}
\vskip -3.3cm 
\caption[l]{\begin{sloppypar}
Energy bounds, $\Omega_C({\bf q})$, for both triplet continua (dashed
dotted) versus triplet--mode energy, $E({\bf q})$, (solid) as well as
combined spectral weight of continua, $W_C({\bf q})$, (dotted) at
$T=0.05$ versus spectral weight of triplet mode, $W_T({\bf q})$,
(dashed) for momenta along $q_b=q_c$. ($b=c=1$, $J=1$ and $t_1,t_2$ as
in fig.~\ref{bobofig6}.)
\end{sloppypar}}
\label{bobofig7}
\end{figure}

\begin{figure}[tb]
\vskip -3.8cm 
\centerline{\psfig{file=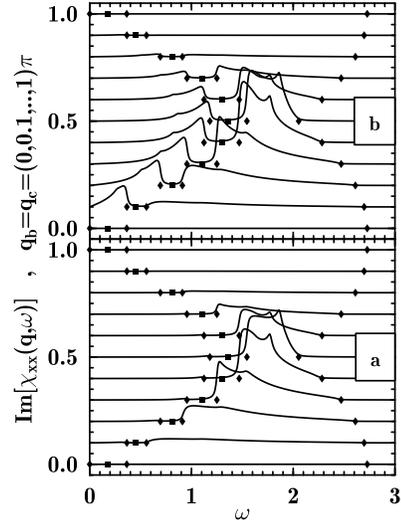,width=9.5cm}}
\vskip -2.3cm 
\caption[l]{\begin{sloppypar}
Triplet continua for various momenta along $q_b=q_c$ and for two
temperatures $T=0.05$ (a) and $T=0.15$ (b). Bottom--most curve
corresponds to ${\bf q}=0$ with $0.1$ incremental y--axis offset
for each consecutive momentum.  Solid diamonds: bounds of triplet
continua. Solid square: triplet--mode energy.  ($b=c=1$, $J=1$ and
$t_1,t_2$ as in fig.~\ref{bobofig6}.)
\end{sloppypar}}
\label{bobofig8}
\end{figure}

\subsection{Dynamic Structure Factor}\label{SqwSection}

Magnetic excitations are observed by measuring the dynamic structure
factor $S({\bf q},\omega)$ which is related by analytic continuation
and the fluctuation dissipation theorem $S({\bf q},\omega)=
\mbox{Im}[\chi({\bf q},\omega)]/(1-e^{-\omega/T})$ to the dynamic
spin susceptibility
\begin{equation}\label{17}
\chi_{\alpha\beta}({\bf q},\tau)= \langle T_\tau [S^\alpha_{\bf
q}(\tau) S^\beta_{\bf q}] \rangle
\;\;\;.\end{equation}
Here ${\bf q}$ is the momentum, $\tau$ the imaginary time, and
$T_\tau$ refers to time ordering.  Since the physical spin is a
composite operator of the bond--bosons the information obtained from
the dynamic susceptibility is not restricted to the triplet dispersion
-- even at the LHP level. This will be clarified in the remainder of
this section.

Within the LHP representation the spin operator in momentum space is
\begin{eqnarray}\label{18}
S^\alpha_{\bf q}&=& \frac{1}{4}(1-e^{iq_c})
(t_{{\bf q}\,\alpha}^{\dagger}+
t_{{\bf -q}\,\alpha}^{\phantom{\dagger}})
\nonumber \\
&& -i\frac{1}{4\sqrt{D}}(1+e^{iq_c})
\sum_{\bf k\in {\cal B}}
\varepsilon_{\alpha\beta\gamma}
t_{{\bf k}+{\bf q}\,\beta}^{\dagger}
t_{{\bf k}\,\gamma}^{\phantom{\dagger}}
\;\;\;.\end{eqnarray}
Two qualitatively different excitations appear on the rhs.: a sharp
triplet mode due to the first term and a continuum of two--triplet
states due to the second.  The momentum dependent form factors $(
1\mp\exp (iq_c)$ lead to a vanishing weight of the the triplet mode
(continuum) at $q_c=0$ ($q_c=\pi$). Since $H_{LHP}$ does not conserve the
bare triplet number, the continuum is divided in between two excitations
of different nature, i.e. virtual excitations at energies $E_{\bf k+q}+
E_{\bf k}$ and real excitations at energies $E_{\bf k+q}- E_{\bf
k}$. While the former are due to ground--state quantum--fluctuations
and occur at all temperatures, the latter result from excitations
across the spin gap and are present only at finite
temperatures. Evaluating the dynamic susceptibility by standard
methods I obtain
\begin{eqnarray}\label{19}
\lefteqn{
\chi_{\alpha\beta}({\bf q},\omega_n) = \delta_{\alpha\beta} \frac{1}{4}
\left\{(\cos(q_c) - 1)\frac{1}{(i\omega_n)^2-E^2_{\bf q}}
\right.}
\nonumber \\ \nonumber \\
&&+(\cos(q_c)+1) \frac{1}{2D} \sum_{{\bf k}\in {\cal B}} \left[
\frac{1+\epsilon_{{\bf k}+{\bf q}}+\epsilon_{\bf k} +
E_{{\bf k}+{\bf q}}E_{\bf k}}{E_{{\bf k}+{\bf q}}E_{\bf k}}\times \right.
\nonumber \\ &&
\phantom{(1+\cos(q_c))\frac{1}{2D}\sum_{{\bf k}\in {\cal B}}
\frac{1+\epsilon_{\bf k}}{E_{{\bf k}+{\bf q}}E_{\bf k}}}
\frac{n(E_{{\bf k}+{\bf q}})-n(E_{\bf k})}
{i\omega_n+E_{{\bf k}+{\bf q}}-E_{\bf k}}
\nonumber \\ \nonumber \\ &&
\phantom{(1}
+\frac{(1+\epsilon_{{\bf k}+{\bf q}}+\epsilon_{\bf k} -
E_{{\bf k}+{\bf q}}E_{\bf k})(E_{{\bf k}+{\bf q}}+E_{\bf k})}
{E_{{\bf k}+{\bf q}}E_{\bf k}}\times
\nonumber \\ &&
\phantom{(1\cos(q_c))\frac{1}{2D}\sum_{{\bf k}\in {\cal B}} [ }
\left.\left. \frac{n(E_{{\bf k}+{\bf q}})+n(E_{\bf k})+1}
{(i\omega_n)^2-(E_{{\bf k}+{\bf q}}+E_{\bf k})^2} \right] \right\}
\;\;\;,\end{eqnarray}
where $\omega_n=2n\pi T$ is a Bose Matsubara--frequency.

As anticipated the dynamical susceptibility (\ref{19}) exhibits finite
spectral intensity at $E_{\bf k}$, $E_{\bf k+q}-E_{\bf k}$, and
$E_{\bf k+q}+E_{\bf k}$. This is summarized in fig.~\ref{bobofig7} which
displays the two continua with respect to the triplet mode along a
particular momentum space direction. Parameters identical to those of
fig.~\ref{bobofig6} have been chosen.  The triplet mode is situated in a
gap between the low--energy continuum due to thermal excitations and that
at high energies due to quantum fluctuations. At the zone center the
latter is gaped by $2 \Delta$.  This is consistent with recent INS
data from the D--phase of CuGeO$_3$ \cite{Ain97a}. In addition to the
magnon these experiments indicate an 'unexpected' high--energy
continuum which is separated from the magnon by an additional gap.
The observed zone--center continuum--to--magnon gap--ratio is
approximately $2$. Deviations from the latter value of $2$
are possibly due to triplet--triplet interactions which are beyond the
LHP approach. Additionally, fig.~\ref{bobofig7} shows the weight of the
triplet mode, as well as that of the combined continua at $T=0.05J$.
At $T\ll J$ the continuum weight is almost completely due to quantum
fluctuations while, independent of temperature, the triplet weight is
given by
\begin{equation}\label{20}
W_T({\bf q})=\frac{\pi}{8} \frac{1-\cos(q_c)}{E_{\bf q}} \,
\stackrel{(\pi,\pi)}{=} \;\frac{\pi}{4\Delta}
\;\;\;,\end{equation}
The leftmost expression refers to the maximum of the triplet
weight. The corresponding momentum, i.e. $(\pi,\pi)$, indicates the
instability towards AMF as $\Delta\rightarrow 0$.

Figure~\ref{bobofig8} depicts the spectral intensity
$\mbox{Im}[\chi_{xx} ({\bf q},\omega_n\rightarrow
-i\omega+\nu)]$ of the continua as a function of frequency for two
temperatures and for various momenta along a direction in reciprocal
space identical to that of fig.~\ref{bobofig7}.  The ${\bf k}$--sums
of (\ref{19}) have been performed numerically on a 600$\times$300
lattice.  In order to obtain sufficient smoothing the frequency has
been shifted off the real axis by $\nu=0.05$.  The parameters
correspond to those of fig.~\ref{bobofig6}.  As a guide to the eye the
position of the triplet mode is labeled by solid squares while solid
diamonds in this figure label the exact locations of the spectral
bounds of the continua. Although smeared due to the imaginary
broadening, van--Hove singularities are clearly observable at the
spectral bounds as well as other characteristic energies within the
continua.  Evidently the quantum fluctuations exhibit largest weight
at intermediate wave vectors while at higher temperatures additional
weight appears at smaller momentum due to thermal excitations.
Even though the intensity in fig.~\ref{bobofig8} decreases
both, as ${\bf q}$ approaches $(0,0)$ and $(\pi,\pi)$, only in the
latter case this is due to the form factor in (\ref{18}) while in the
former case this is a consequence of the conservation of the total spin.

\section{Conclusion}

In summary I have studied static and dynamic properties of a
frustrated and dimerized 2D quantum spin--model using the
bond--operator method.  The ground state energy and the spin gap have
been gauged against known results from the 1D limiting cases of this
model. Effects of dimerization are found to be described almost
quantitatively while the influence of frustration is captured
qualitatively.  The dynamic structure factor has been analyzed
and displays two characteristic features, i.e. a well defined magnon
excitation and a temperature dependent continuum. The magnon
dispersion shows a characteristic lifting of degeneracies, different
from purely one--dimensional models, and agrees very well with INS
data on CuGeO$_3$. The low--temperature continuum exhibits a
zone--center gap twice that of the magnon. This is also consistent
with INS experiments on CuGeO$_3$.

\acknowledgements

I am grateful to P. Fulde and the Max--Planck--Institut f\"ur Physik
komplexer Systeme for their kind hospitality. It is a pleasure to
thank G. Uhrig for helpful comments and for communicating his results
prior to publication. Stimulating discussions with B. B\"uchner and
E. M\"uller--Hartmann are acknowledged. This work has been supported
in part by the Deutsche Forschungsgemeinschaft through the SFB 341.

\begin{appendix}

\section{Bond--Operator Mean--Field Theory (MFT)}\label{A}

An alternative approach to the Hamiltonian (\ref{5}) arises by
introduction of a set of local Lagrange multipliers $\eta_{\bf l}$ to
enforce the constraint (\ref{3})
\begin{equation}\label{A1}
\tilde{H}=H - \sum_{{\bf l}\in{\cal D}}
\eta_{\bf l}(s_{{\bf l}}^{\dagger}
s_{{\bf l}}^{\phantom{\dagger}}
+t_{{\bf l}\,\alpha}^{\dagger}
t_{{\bf l}\,\alpha}^{\phantom{\dagger}}-1)
\;\;\;.\end{equation}
To treat this Hamiltonian one replaces the local constraint by a
global one, i.e. $\eta_{\bf l}=\eta$, and introduces a mean--field
(MF) decoupling of all quartic terms leading to an effective quadratic
Hamiltonian \cite{Sachdev90a}. This Hamiltonian has an overall {\em
negative} prefactor to the $s_{{\bf l}}^{\dagger} s_{{\bf l}}^{
\phantom{\dagger}}$--term which implies Bose condensation of the
singlets.  Therefore $s_{{\bf l}}^{\dagger}=s_{{\bf
l}}^{\phantom{\dagger}}=\langle s_{{\bf l}}^{\phantom{
\dagger}}\rangle=s$ is assumed. Moreover it can be shown that
contributions from the triplic and quartic triplet--terms $H_2$ and
$H_3$ to the MFT can be neglected \cite{GopalanXXa,Brenig97b}.  The MF
Hamiltonian reads
\begin{eqnarray}\label{10}
\lefteqn{
H_{MFT}=D(-\frac{3}{8}-\frac{3}{4}s^2-\eta s^2+\frac{5}{2}\eta)
}
\\
&&+\frac{1}{2}\;\sum_{{\bf k}\in{\cal B}}
\Psi_{{\bf k}\,\alpha}^\dagger
\left[\begin{array}{cc}
\frac{1}{4}-\eta+s^2\epsilon_{\bf k} & s^2\epsilon_{\bf k} \\
s^2\epsilon_{\bf k} & \frac{1}{4}-\eta+s^2\epsilon_{\bf k}
\end{array}\right]
\Psi_{{\bf k}\,\alpha}^{\phantom{\dagger}}
\nonumber
\;\;\;,\end{eqnarray}
with notations equivalent to (\ref{7}). Analogous to the latter equation
(\ref{10}) represents three dispersive triplet excitations, however,
with a modified dispersion relation. After a Bogoliubov transformation
one gets \cite{etag0}
\begin{eqnarray}\label{11}
H_{MFT}=&&D(-\frac{3}{8}-\frac{3}{4}s^2-\eta s^2+\frac{5}{2}\eta)
\nonumber \\
&&+\sum_{{\bf k}\in{\cal B},\alpha}
E^{MFT}_{\bf k}(b_{{\bf k}\,\alpha}^{\dagger}
b_{{\bf k}\,\alpha}^{\phantom{\dagger}}
+\frac{1}{2})
\\
\nonumber \\ \label{11a}
E^{MFT}_{\bf k}=&&(\frac{1}{4}-\eta)\sqrt{1+d \epsilon_{\bf k}}
\;\;\;,\end{eqnarray}
where $d=2s^2/(1/4-\eta)$. The $b_{{\bf k}\, \alpha}$ quasi--particles
follow from expressions identical to (\ref{9}) with $\epsilon_{\bf k}$,
$1+\epsilon_{\bf k}$, and $E_{\bf k}$ replaced by $s^2\epsilon_{\bf
k}$, $\frac{1}{4}-\eta+s^2\epsilon_{\bf k}$, and $E^{MFT}_{\bf k}$,
respectively.  The Lagrange multiplier and the singlet amplitude have
to be determined by solving the saddle--point equations
$\langle\partial H_{MFT}/\partial\eta\rangle=0$ and $\langle\partial
H_{MFT}/\partial s\rangle=0$. This leads to
\begin{eqnarray}\label{12a}
0&=&s^2-\frac{2}{5}+\frac{3}{D}\sum_{{\bf k}\in{\cal B}}
\frac{1+d\epsilon_{\bf k}/2}{\sqrt{1+d \epsilon_{\bf k}}}
(\langle b_{{\bf k}\,x}^{\dagger}
b_{{\bf k}\,x}^{\phantom{\dagger}}\rangle +\frac{1}{2})
\\ \nonumber \\ \label{12b}
0&=&\frac{3}{4}+\eta-\frac{3}{D}\sum_{{\bf k}\in{\cal B}}
\frac{\epsilon_{\bf k}}{\sqrt{1+d \epsilon_{\bf k}}}
(\langle b_{{\bf k}\,x}^{\dagger}
b_{{\bf k}\,x}^{\phantom{\dagger}}\rangle +\frac{1}{2})
\;\;\;.\end{eqnarray}
These selfconsistency equations can be solved by combining them into a
single one for the variable $d$ only
\begin{eqnarray}\label{13a}
d&=&5-\frac{3}{D}\sum_{{\bf k}\in{\cal B}}
\frac{2}{\sqrt{1+d \epsilon_{\bf k}}} ( \langle b_{{\bf k}\,x}^{\dagger}
b_{{\bf k}\,x}^{\phantom{\dagger}}\rangle+\frac{1}{2} )
\\ \label{13b}
&\stackrel{(T=0)}{=}&
5-\frac{3}{D}\sum_{{\bf k}\in{\cal B}}
\frac{1}{\sqrt{1+d \epsilon_{\bf k}}}
\;\;\;,\end{eqnarray}
where $\eta$ follows by insertion of $d$ into (\ref{12b}). This
completes the description of the bond--operator MFT.

\end{appendix}

\end{document}